\documentstyle[aps,preprint,prd]{revtex}

\begin{document}

\draft

\title{On ferrimagnetic phases in chiral Yukawa models}

\author{J.L.\ Alonso}
\address{Departamento de F\'\i sica Te\'orica,
           Universidad de Zaragoza, Facultad de Ciencias,
           50009 Zaragoza, Spain}

\author{Ph.\ Boucaud}
\address{Laboratoire de Physique Th\'eorique et Hautes
           Energies, Universit\'e de Paris XI, 91405 Orsay Cedex,
           France
           (Laboratoire associ\'e au CNRS) }

\author{A.J.\ van der Sijs\footnote{email: arjan@sol.unizar.es.}}
\address{Departamento de F\'\i sica Te\'orica,
           Universidad de Zaragoza, Facultad de Ciencias,
           50009 Zaragoza, Spain}

\preprint{Preprint\ \ Zaragoza DFTUZ/95/07,\ \ LPTHE Orsay-95/13,\ \
hep-lat/9503006}

\date{February 1995}

\maketitle

\begin{abstract}
We discuss the phase structure of chiral Yukawa
models in the mean-field approximation.
In particular, we examine under which conditions a ferrimagnetic phase
appears, by calculating the slopes of possible second order phase transition
lines near a critical point.
Our results contrast with some statements which appeared in the literature
recently.
\end{abstract}

\pacs{PACS numbers: 11.15.Ha, 05.70.Fh, 14.80.Bn}

\narrowtext

\section{Introduction}
\label{sec1}

Chiral Yukawa models on the lattice
are interacting fermion-Higgs field theories
with a chiral symmetry, which are useful for studying the Standard Model
on the lattice in the absence of gauge couplings
\cite{Shige-lat}.
The phase structure of such models, in particular
the location of fixed points and second order phase transition lines, is
of interest for the definition of corresponding continuum field theories.

The action of a chiral Yukawa model is of the generic form
\begin{eqnarray}
S &=& -\frac k2 \sum_{x,\mu} \mbox{Tr}\  [\Phi_x^+ \Phi_{x+\hat\mu}
    + \mbox{h.c.} ]
        \nonumber \\
  &&\mbox{}+ y \sum_x \overline\Psi_x (P_L \Phi_x^+
     + P_R \Phi_x) \Psi_x
\label{S} \\
  &&\mbox{+ fermion kinetic term + Higgs potential} , \nonumber
\end{eqnarray}
the detailed definition of these terms varying from one model to the
other.
With `triviality' in mind, we take the bare Higgs
self-coupling equal to $\infty$, which fixes the radial mode of the
Higgs field $\Phi$.
The phase diagram also depends on the details of the model, but the
phase structure in the small to intermediate-$y$ region is similar
in most models.
We shall concentrate on this part of the phase diagram (see Fig.~\ref{fig1}),
which is sufficient for our purposes, but the considerations
presented here are also valid for critical points at large-$y$,
where applicable.
There are paramagnetic (PM), ferromagnetic (FM), antiferromagnetic (AM)
and possibly ferrimagnetic (FI) phases, characterized by the vacuum
expectation values of the spatial average and staggered average (this
means that fields at odd sites contribute with a minus sign) of
the Higgs field and the fermion condensate.
In the context of chiral Yukawa models, phases where both the Higgs field
and its staggered analogue have non-zero expectation value
are known as FI phases \cite{Shigemitsu}.

In recent years, various kinds of Yukawa models, based on different
lattice fermion actions and with different symmetry groups, have been
studied both using mean-field methods and numerically \cite{Shige-lat}.
Here we shall concentrate on the mean-field approach \cite{MF}.
Because of the difficulty of dealing with the fermion determinant exactly,
mean-field techniques are often applied in combination with small or
large-$y$ expansions.
In Ref.\ \cite{Zaraphase} this type of analysis was extended to FI phases.
Although FI phases start off at values of $y$ where one might question the
validity of the small-$y$ expansion involved, it was argued that
the real expansion parameter is the quantity
$y\langle\overline\Psi\Psi\rangle$, which remains small as long as one keeps
close to the PM phase.
Furthermore, the phase structure of the models studied, obtained from
numerical simulation, appeared to be well-described by the mean-field results.

In a recent publication \cite{tomzen},
mean-field calculations were presented for a
general class of chiral Yukawa models with different symmetry groups
and lattice fermion actions.
(The details of this approach differ somewhat from other
applications of mean-field methods in the truncation of the fermion
determinant, but this is of no importance here.)
The authors claim that, in general,
if the PM--FM and PM--AM phase transition lines
intersect at a point $A$ (cf.\ Fig.~\ref{fig1}),
then they must necessarily continue beyond the point $A$ with smooth first
derivatives at $A$.  There would always be
a FI phase, with the PM--FM line continuing as an AM--FI phase transition
line and the PM--AM line as an FM--FI line.
(Everything within the mean-field approximation.)
This is in contrast with the results of Refs.\ \cite{Zaraphase,ZaraphaseR},
where discontinuities were observed in the slopes of the phase transition
lines at $A$, with indications in one model \cite{ZaraphaseR} that a FI
phase is absent at all. (Again, within the mean-field approximation.)

These discrepancies motivated us to carry out the present study,
which may however have a wider applicability.
We analyse the phase structure around the point $A$ from a general
point of view, in the mean-field approximation.
It is found that
the first derivatives of the second order phase transition lines,
assuming that they intersect, are in general {\em discontinuous\/}
at $A$, as in Fig.~\ref{fig1}, and no conclusion can
be drawn {\em a priori\/} about the existence of a FI phase.
After presenting a general demonstration in Sec.\ \ref{sec2}, we
illustrate the results with a simple example in Sec.\ \ref{sec3}.
For definiteness we keep in mind a phase structure as in
Fig.~\ref{fig1}.

\section{Phase structure at $A$}
\label{sec2}

In the mean-field approximation,
the phase of the system at a point $(y,k)$ is determined by
minimizing the free energy $F$ with respect to a number of mean fields
$h^i$ and staggered mean fields $h_s^i$ $(i=1,\ldots,N)$,
collectively denoted as $h$, $h_s$.
($F$ is a function of $h$ and $h_s$ with coefficients depending on $y$ and
$k$.)
This gives the mean-field equations
\begin{equation}
\frac{\partial F}{\partial h} \ =\  0,\ \ \ \ \ \ \
\frac{\partial F}{\partial h_s} \ =\  0.
\label{mfeqs}
\end{equation}
A solution of these equations corresponds to a (local) minimum
of $F$ if the matrix $F''$
of second derivatives at the solution is positive definite.
A second order phase transition occurs when one of the eigenvalues of
this matrix goes through zero: a negative mode develops, destabilizing
the original solution and replacing it by one with a lower free energy,
belonging to a different phase.
Therefore, the condition for a second order phase transition line is
given by the additional equation
\begin{equation}
\det F'' \ =\  0.
\label{detzero}
\end{equation}
If there is only one mean field $h$ and one staggered mean field $h_s$
this becomes\footnote{There is a square missing in the corresponding
formula (17) of Ref.~\cite{tomzen}.}
\begin{equation}
\frac{\partial^2 F}{\partial h^2} \
\frac{\partial^2 F}{\partial h_s^2} \ -\
\left( \frac{\partial^2 F}{\partial h \; \partial h_s} \right)^2 \ =\  0.
\label{detzero1}
\end{equation}

In the chiral Yukawa models under consideration, the free energy $F$
is symmetric under a sign change of all the
fields $h^i$ at the same time or all the fields $h_s^i$ at the same time.
This implies the absence of terms containing odd powers of $h$ or $h_s$ in
the Taylor expansion of $F$, such as $h h_s$.
Furthermore, at the four phase transition lines under study here,
either all the $h^i=0$ or all the $h_s^i=0$ or both.
Hence $\partial^2 F/\partial h^i \partial h_s^j = 0$ there
and Eq.\ (\ref{detzero}) simplifies,
\begin{equation}
\det \left( \frac{\partial^2 F}{\partial h^2} \right) \
\det \left( \frac{\partial^2 F}{\partial h_s^2} \right) \ =\  0.
\label{detzero2}
\end{equation}
Which of the two factors in this expression is zero depends on which
phase transition line one considers; at the intersection point $A$
both factors are zero.

For definiteness we shall focus on the PM--FM and AM--FI transitions.
Then condition (\ref{detzero2}) becomes
\begin{equation}
\det\frac{\partial^2 F}{\partial h^2} (h=0, h_s) \ =\  0,
\label{detzero3}
\end{equation}
since in this case the unstable mode is in the direction of the $h$ fields.
Along the PM--FM line, $h_s=0$ and the solution to Eq.\ (\ref{detzero3})
is readily obtained.
Along the AM-FI line, however, $h_s\neq 0$. One has to solve
Eqs.\ (\ref{mfeqs})
and (\ref{detzero3}) simultaneously,
to find $h_s(y,k)$ and the function $k_c(y)$ parametrizing the line.
This procedure was applied numerically for phase diagrams of chiral Yukawa
models based on the Zaragoza proposal \cite{Zaraprop} with the `most local'
and `Roma I' \cite{RomaI} fermion actions \cite{Zaraphase,ZaraphaseR}.

Here, we are especially interested in
the behaviour of the free energy $F$
in the vicinity of the point $A$.  In this region, close to the PM phase,
the mean fields are
small and one can study the phase structure by considering an expansion
in $h$ and $h_s$.
In Ref.~\cite{tomzen} it is claimed that it is
sufficient for this purpose
to know $F$ up to quadratic terms in $h$ and $h_s$.
The following calculation of the derivatives of the phase transition
lines at $A$ will show, however, that quartic terms
(or, in their absence, higher order terms) are indispensable;
neglecting them leads to incorrect statements
about a possible FI phase near $A$.

The PM--FM and AM--FI phase transition lines are given by a continuous
function $k_c(y)$ whose derivative we wish to determine.
Defining
\begin{equation}
f \ \equiv \
\det\left.\frac{\partial^2 F}{\partial h^2}\right|_{h=0},
\label{fdef}
\end{equation}
we can write Eq.\ (\ref{detzero3}) for these lines as
\begin{equation}
f(y,k_c(y),h_s^2(y,k_c(y))) = 0.
\label{fcond}
\end{equation}
Taking the derivative of this equation in the direction tangential to
the line we find
\begin{equation}
0 = \frac{df}{dy} = \frac{\partial f}{\partial y}
 + \frac{\partial f}{\partial k} \frac{dk_c}{dy}
 + \frac{\partial f}{\partial h^2_s} \left(
 \frac{\partial h^2_s}{\partial y}
 + \frac{\partial h^2_s}{\partial k} \frac{dk_c}{dy} \right),
\label{dfcond}
\end{equation}
leading to a slope
\begin{equation}
\frac{dk_c}{dy} = -\left(
 \frac{\partial f}{\partial y}
 + \frac{\partial f}{\partial h^2_s}
 \frac{\partial h^2_s}{\partial y} \right) \left/ \left(
 \frac{\partial f}{\partial k}
 + \frac{\partial f}{\partial h^2_s}
 \frac{\partial h^2_s}{\partial k} \right) \right. .
\label{slope}
\end{equation}
On the left hand (PM--FM) side of point $A$, $h^2_s=0$
and Eq.\ (\ref{slope}) reduces to
\begin{equation}
\left( \frac{dk_c}{dy} \right)_{\rm PM-FM} \ =\
 -\frac{\partial f}{\partial y} \left/
 \frac{\partial f}{\partial k} \right. .
\label{slope2}
\end{equation}
On the AM--FI side, however, it is reasonable to
assume that $h_s^2$ approaches $A$ linearly in
$y-y_A$ and $k-k_A$, in accordance with a mean-field critical exponent
of $1/2$ for $h_s$ in this region
(cf.\ the example in Sec.\ \ref{sec3}).
It follows that both the numerator and the denominator of Eq.\ (\ref{slope})
receive an additional non-zero contribution on this side, so that
$dk_c/dy$ is discontinuous at $A$.

A similar analysis can be carried out for the change of slope between
the PM--AM and FM--FI lines.
We emphasize, however, that the AM--FI and PM--FI lines thus obtained
(at least, infinitesimally close to $A$)
remain `candidate phase transition lines' only until free energy
considerations establish that the phases on both sides of the lines
correspond to absolute minima of the free energy.
The slope of the AM--FI line suggested by these calculations may,
for example, come out bigger than that of the PM--FI line.
(This does in fact happen, at least to lowest order in the Yukawa
coupling $y$, in a chiral Yukawa model based
on the Roma I action \cite{ZaraphaseR}.)
In that case,
a comparison of free energy values should indicate which
of the calculated lines does or do not correspond to a true second order
phase transition.

\section{A simple example}
\label{sec3}

We would like to illustrate these results with a simple model of
one mean field $h$ with its staggered analogue $h_s$,
described by a quartic free energy $F$,
\begin{equation}
F = -\frac{1}{2} a h^2 - \frac{1}{2} b h_s^2 + \frac{1}{4} c h^2 h_s^2
  + \frac1{24} d h^4 + \frac1{24} e h_s^4 ,
\label{Fexp}
\end{equation}
where the parameters $a,\ldots,e$ are well-behaved
functions of $y$ and $k$.
We assume the presence of PM, FM and AM phases as in Fig.~\ref{fig1},
{\em i.e.}, $a$ increases and
$b$ decreases with increasing $k$, and $a(y,k)$ and $b(y,k)$
are such that the PM--FM and PM--AM lines meet in a point $A$.
Furthermore, for stability of $F$, we require $d>0$, $e>0$, $c>-\sqrt{de}/3$.
Apart from these stability conditions, one can consider this model as the
expansion of a general free energy in the neighbourhood of the point $A$
up to quartic terms in the fields.

In those points of the $(y,k)$-plane where
both $a<0$ and $b<0$, the mean-field equations (\ref{mfeqs}) imply that
the system is in a PM phase with $h=h_s=0$ and free energy normalized to
zero.
If $a>0$ or $b>0$ we are in one of the broken phases, FM, AM or FI.
Condition (\ref{detzero3}) for the PM--FM and AM--FI second order
phase transition lines becomes
\begin{equation}
-a + \frac{1}{2} c h_s^2 = 0.
\label{detzero4}
\end{equation}
Along the PM--FM line this becomes simply $a=0$, whereas along the AM--FI
line we need the value of $h_s$ which follows from the mean-field equations
(\ref{mfeqs}), taken at $h=0$,
\begin{equation}
-b h_s + \frac16 e h_s^3 = 0.
\label{hseq}
\end{equation}
At the point $A$,
both $a=0$ and $b=0$,
whereas in general $c$, $d$
and $e$ are non-zero. Close to $A$ we can therefore write,
\begin{eqnarray}
a(y,k) &\ =\ & a_y (y-y_A) + a_k (k - k_A) + h.o., \nonumber \\
b(y,k) &\ =\ & - b_y (y-y_A) - b_k (k - k_A) + h.o., \nonumber \\
c(y,k) &\ =\ & c_0 + h.o., \nonumber \\
d(y,k) &\ =\ & d_0 + h.o., \nonumber \\
e(y,k) &\ =\ & e_0 + h.o.,
\label{abcde-exp}
\end{eqnarray}
where $h.o.$ stands for higher orders in the $y-y_A,k-k_A$ expansion.
The signs in front of $a_k$ and $b_k$ have been chosen such that both
are positive.
In terms of these coefficients, the slopes of the PM--FM and PM--AM lines
at $A$ are
(cf.\ Eq.\ (\ref{slope2}))
\begin{eqnarray}
R_{\rm PM-FM}
  &\ \equiv\ &\left( \frac{dk_c}{dy} (A) \right)_{\rm PM-FM}
  \ =\  -\frac{a_y}{a_k},
\label{rFMPM} \\
R_{\rm PM-AM} &\ =\ & -\frac{b_y}{b_k}.
\label{rAMPM}
\end{eqnarray}
In order to find the AM--FI slope at $A$ we solve
Eqs.\ (\ref{detzero4}) and (\ref{hseq}) to lowest order in $y-y_A$ and $k-k_A$.
We find
\begin{equation}
h_s^2 = -\frac6{e_0} (b_y (y-y_A) + b_k (k-k_A)) + h.o.,
\label{hssol}
\end{equation}
corresponding to a critical exponent $1/2$ for $h_s$
as mentioned earlier,
and
\begin{equation}
R_{\rm AM-FI} = -\frac{a_y + 3(c_0/e_0) b_y}{a_k + 3(c_0/e_0) b_k}
\label{rAMFI}
\end{equation}
(cf.\ Eq.\ (\ref{slope})).

It is interesting to express Eq.\ (\ref{rAMFI}) as a `weighted average' of
the PM--FM and PM--AM slopes:
\begin{equation}
R_{\rm AM-FI} = \frac{e_0 a_k R_{\rm PM-FM} + 3 c_0 b_k
     R_{\rm PM-AM}}{e_0 a_k + 3 c_0 b_k}.
\label{rAMFI2}
\end{equation}
Similarly, we find for the FM--FI slope
\begin{equation}
R_{\rm FM-FI} = \frac{3c_0 a_k R_{\rm PM-FM} + d_0 b_k
     R_{\rm PM-AM}}{3c_0 a_k + d_0 b_k}.
\label{rFMFI2}
\end{equation}

The discontinuities in the slopes are evident.
Only for $c=0$ the first derivatives of these phase transition lines are
continuous (ignoring special cases like $c\neq 0, c_0=0$).
This corresponds to the trivial case that there is no coupling
between $h$ and $h_s$, as mentioned earlier.
Note that in the limit $e_0 \rightarrow 0$ Eq.\ (\ref{rAMFI2}) tends
to $R_{\rm AM-FI} =
R_{\rm PM-AM}$, while in the limit $d_0 \rightarrow 0$ which can be taken
simultaneously Eq.\ (\ref{rFMFI2}) leads to $R_{\rm FM-FI} = R_{\rm PM-FM}$.
In this case the AM--FI and FM--FI lines would be `interchanged'
compared with the $c=0$ case.
This obviously indicates that at least one of the calculated `candidate
transition lines' does not represent a genuine phase transition, and
additional free energy considerations must determine the real
nature of the transitions, as discussed before.

It is also instructive to consider the AM--FI and FM--FI slopes as a
function of $c_0$.
In the limit that $c_0$ approaches its `stability lower bound'
$-\sqrt{d_0 e_0}/3$, these lines form a 180 degree angle.  Upon
increasing $c_0$ this angle decreases, until it vanishes
for $c_0 = \sqrt{d_0 e_0}/3$.  For still larger values of $c_0$,
corresponding to strong coupling between $h$ and $h_s$ in the free energy,
the FI phase disappears (at least in a neighbourhood of the point $A$;
higher order terms in Eqs.\ (\ref{abcde-exp}) may give rise to a FI phase
a little farther out, as in Fig.~\ref{fig2}), and instead there is a
first order transition separating the AM and FM phases.

\section{Conclusion}
\label{sec4}

For determining the phase structure close to the point $A$ where several
phases meet, in a mean-field approximation, it is not sufficient to
expand the free energy up to terms quadratic in the mean fields.
Quartic terms are crucial for determining the slopes of
the transition lines enclosing the FI phase, and in general the slopes
are discontinuous at $A$.
Such behaviour was observed in mean-field studies
of phase diagrams of chiral Yukawa models \cite{Zaraphase,ZaraphaseR},
where the (candidate) transition
lines bordering the FI phase were determined numerically.

These results explain the discrepancy signalled in Sec.\ \ref{sec1}.
The authors of Ref.~\cite{tomzen} incorrectly assumed that it is sufficient
to consider the free energy up to quadratic terms in the mean fields.
As a consequence, the conclusions drawn there about the presence of
FI phases and the
transition lines separating them from the FM and AM phases are incorrect.
However, the results for the PM--FM and PM--AM lines are not affected.
We emphasize that these conclusions do not depend on the approximations
made in the treatment of the fermion determinant in chiral Yukawa models.
The actual location and slopes of the various phase transition lines do,
however, because different approximations lead to different
`effective' free energies.
Hence, improved approximations of the fermion determinant may change
the conclusion about the existence of a FI phase, apart from the
limitations of the mean-field approximation.\\

This work was supported by EC contracts ERBCHBICT941067 and
CHRX-CT92-0051, by DGICYT (Spain) and by Acci\'on Integrada
Hispano-francesa HF94-150B.

\begin{figure}
\caption{
Generic phase diagram of a chiral Yukawa model in the small-$y$ region.
On the vertical axis is the hopping parameter $k$ of the Higgs field,
on the horizontal axis is the Yukawa coupling $y$ between the Higgs field
and the fermions $\Psi$.}
\label{fig1}
\end{figure}

\begin{figure}
\caption{
Possible (mean-field) phase structure for $c_0 > \protect\sqrt{d_0e_0}/3$,
with
a first order AM--FM transition (dashed line) and a FI phase away from
the point $A$.}
\label{fig2}
\end{figure}

\end{document}